\documentclass[useAMS,usenatbib]{mn2e}

\usepackage{graphicx}
\usepackage{psfig}

\newcommand\aj{AJ}
\newcommand\apj{ApJ}
\newcommand\aap{A$\&$A}
\newcommand\mnras{MNRAS}
\newcommand\pasp{PASP}

\newcommand{\bm}[1]{{\mbox{\boldmath $#1$}}}

\title[Scaling relations from photo-$z$s]
       {Convolution and deconvolution based estimates of 
        galaxy scaling relations from photometric redshift surveys}

\author[R. K. Sheth \& G. Rossi]
{Ravi K. Sheth$^1$\thanks{Email: shethrk@physics.upenn.edu} \& 
 Graziano Rossi$^{2}$\thanks{Email: graziano@kias.re.kr}\\ \\
$^{1}$ Center for Particle Cosmology, University of Pennsylvania, 
       209 South $33^{rd}$ Street, Philadelphia, PA 19104, USA\\
$^{2}$ Korea Institute for Advanced Study, Hoegiro 87, 
       Dongdaemun-Gu, Seoul $130-722$, Korea}
\date{Accepted 2009 December 23. Received 2009 December 21; in original form 2009 October 7}

\pagerange{\pageref{firstpage}--\pageref{lastpage}}
\pubyear{2009}

\begin{document}

\maketitle

\label{firstpage}


\begin{abstract}
In addition to the maximum likelihood approach, there are two other 
methods which are commonly used to reconstruct the true redshift 
distribution from photometric redshift datasets: one uses a 
deconvolution method, and the other a convolution. We show how 
these two techniques are related, and how this relationship can be 
extended to include the study of galaxy scaling relations in photometric 
datasets. We then show what additional information photometric redshift 
algorithms must output so that they too can be used to study galaxy 
scaling relations, rather than just redshift distributions. We also 
argue that the convolution based approach may permit a more efficient 
selection of the objects for which calibration spectra are required. 
\end{abstract}




\section{Introduction}
The next generation of sky surveys will provide reasonably 
accurate photometric redshift estimates, so there is considerable 
interest in the development of techniques which can use these 
noisy distance estimates to provide unbiased estimates of galaxy 
scaling relations.  
While there exist a number of methods for 
estimating photometric redshifts (Budavari 2009 and references therein), 
there are fewer for using these to estimate accurate redshift 
distributions (Padmanabhan et al. 2005; Sheth 2007; Lima et al. 2008;
Cunha et al. 2009), 
the luminosity function (Sheth 2007), or the joint luminosity-size, 
color-magnitude, etc. relations (Rossi \& Sheth 2008; 
Christlein et al. 2009; Rossi et al. 2010).  

Ideally, the output from a photometric redshift estimator is a 
normalized likelihood function which gives the probability that 
the true redshift is $z$ given the observed colors (i.e. Bolzonella et
al. 2000; Collister \& Lahav 2004; Cunha et al. 2009).  
Let ${\cal L}(z|{\bm c})$ denote this quantity; it may be skewed, 
bimodal, or more generally it may assume any arbitrary shape.

Let $\zeta$ denote the mean or the most probable value of this 
distribution (it does not matter which, although some of the logic 
which follows is more transparent if $\zeta$ denotes the mean).  
Often, $\zeta$ (sometimes with an estimate of the uncertainty on 
its value) is the only quantity which is available.  
Therefore, in Section~\ref{dndz} we first consider how $\zeta$ 
compares with the true redshift $z$, and contrast the convolution 
and deconvolution methods for estimating ${\rm d}N/{\rm d}z$ -- while
in Section \ref{cfc} we describe how to reconstruct the 
redshift distribution directly from colors.
Section~\ref{pdf} shows what this implies if one wishes to use 
the full distribution ${\cal L}(z|{\bm c})$.  
Section~\ref{phil} shows how to extend the logic to the luminosity 
function, and Section~\ref{phix} to scaling relations, again by 
contrasting the convolution and deconvolution methods, and showing 
what generalization of ${\cal L}(z|{\bm c})$ is required from the 
photometric redshift codes if one wishes to do this.
A final section summarizes our results.  

Where necessary, we write the Hubble constant as 
$H_0 = 100h~{\rm km~s}^{-1}~{\rm Mpc}^{-1}$, and we assume a 
spatially flat cosmological model with 
$(\Omega_M,\Omega_{\Lambda}, h)=(0.3, 0.7, 0.7)$, where $\Omega_M$ and
$\Omega_{\Lambda}$ are the present-day densities of matter and
cosmological constant scaled to the critical density. 


\section{To convolve or deconvolve?}

\begin{figure*}
\centering
\includegraphics[width=1.0\hsize]{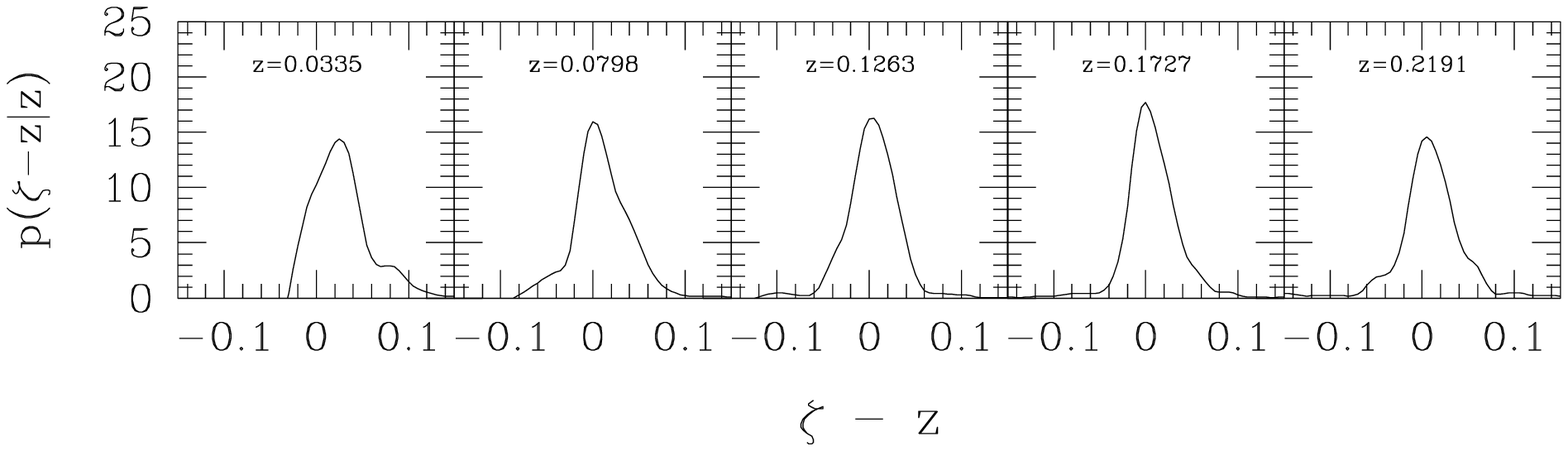}
\includegraphics[width=1.0\hsize]{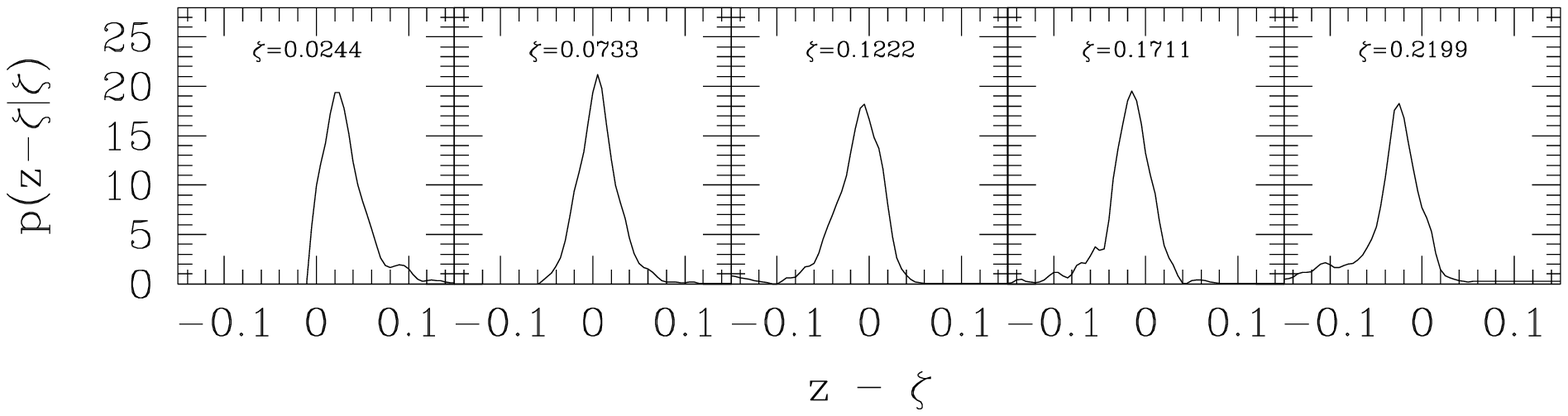}
\caption{Distribution of the difference between spectroscopic 
         and photometric redshifts ($z$ and $\zeta$), at fixed 
         $z$ (top) and $\zeta$ (bottom), in the SDSS early-type 
         galaxy sample.  Note that $p(\zeta|z)$ is rather well 
         centered on $z$, whereas $p(z|\zeta)$ is not centered on $\zeta$.}   
\label{pzzeta}
\end{figure*}

In what follows, we will use spectroscopic and photometric redshifts 
from the SDSS to illustrate some of our arguments. Details of how 
the early-type galaxy sample was selected are in Rossi et al. (2010);  
the photo-$z$s for this sample are from Csabai et al. (2003).  

\subsection{The redshift distribution}
\label{dndz}
Suppose that the true redshifts $z$ are available for a subset of the 
objects; for now, assume that the subset is a random subsample of 
the objects in a magnitude limited catalog.  Ideally, this subset 
would have the same geometry as the full survey, as cross-correlating 
the objects with spectra and those without allows the use of other 
methods (e.g. Caler et al. 2009).  In practice, this may be difficult 
to achieve -- and this is not required for the analysis which 
follows, provided that the photometric redshift estimator does 
not have spatially dependent biases (e.g., as a result of 
photometric calibrations varying across the survey).  

For the objects with spectroscopic redshifts, one can study the 
joint distribution of $\zeta$ and $z$ (see Figure \ref{pzzeta}).  
Typically, most photometric
redshift codes are constructed to return 
 $\langle\zeta |z\rangle \approx z$.  The codes which do so 
are sometimes said to be unbiased, but they are not perfect:
the scatter around the unbiased mean is of order 
 $\sigma_{\zeta|z} \approx 0.05\,(1+z)$.  This scatter, combined 
with the fact that $\langle\zeta |z\rangle \approx z$ means that 
 $\langle z|\zeta\rangle \ne \zeta$:  
the fact that $\langle z|\zeta\rangle$ is {\em guaranteed} to be 
biased is not widely appreciated.  
However, we show below that it matters little whether 
$\langle\zeta |z\rangle$ or $\langle z|\zeta\rangle$
are unbiased -- what matters is that the bias is accurately 
quantified.  

\begin{figure*}
\centering
\includegraphics[width=0.88\hsize]{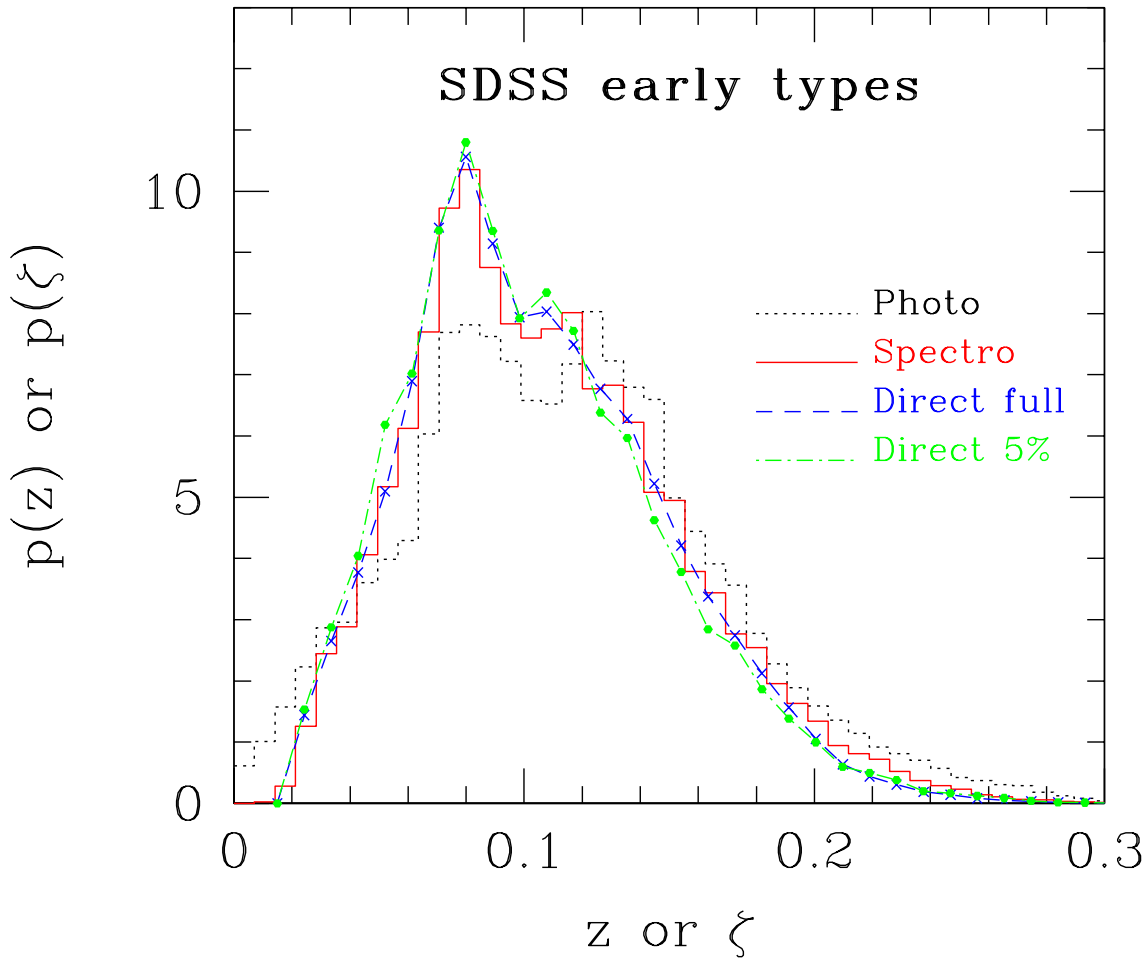}
\caption{Distribution of ${\rm d}{\cal N}/{\rm d}\zeta$ (dotted) 
         and ${\rm d}N/{\rm d}z$ (solid); crosses show the result 
         of convolving ${\rm d}{\cal N}/{\rm d}\zeta$ with $p(z|\zeta)$ 
         (from the bottom panel of Figure~\ref{pzzeta}).}
\label{Nzconv}
\end{figure*}

In particular, if ${\rm d}{\cal N}/{\rm d}\zeta$ and ${\rm d}N/{\rm d}z$ 
denote the distribution of $\zeta$ and $z$ values in the subset of 
the data where both $z$ and $\zeta$ are available, then what matters 
is that $p(\zeta|z)$ and $p(z|\zeta)$, where 
\begin{equation}
 \frac{{\rm d}N(z,\zeta)}{{\rm d}z\,{\rm d}\zeta} 
  = \frac{{\rm d}N(z)}{{\rm d}z}\, p(\zeta|z) 
  = \frac{{\rm d}{\cal N}(\zeta)}{{\rm d}\zeta}\,p(z|\zeta),
\end{equation}
are known.  Note that  
\begin{equation}
 \frac{{\rm d}{\cal N}(\zeta)}{{\rm d}\zeta} 
  \equiv \int {\rm d}z\,\frac{{\rm d}N(z)}{{\rm d}z}\, p(\zeta|z) .
 \label{Nzeta}
\end{equation}
The algorithm in Sheth (2007) assumes that $p(\zeta|z)$, 
measured in the subset for which both $z$ and $\zeta$ are available, 
also applies to the full sample for which $z$ is not available.  
Since ${\rm d}{\cal N}/{\rm d}\zeta$ is measured in the full dataset, 
and $p(\zeta|z)$ is known, a deconvolution is then used to estimate 
the true ${\rm d}N/{\rm d}z$.  

Suppose, however, that one measured $p(z|\zeta)$ instead.  
Then, because 
\begin{equation}
 \frac{{\rm d}N(z)}{{\rm d}z} 
  \equiv \int {\rm d}\zeta\,\frac{{\rm d}{\cal N}(\zeta)}{{\rm d}\zeta}\,
              p(z|\zeta) ,
 \label{Nz}
\end{equation}
one could estimate the quantity on the left hand side by 
`convolving' the two measurables on the right hand side.  
For the data-subset in which both $z$ and $\zeta$ are available, 
this is correct by definition.  Clearly, to use this method on 
the larger dataset for which only $\zeta$ is available, one 
must assume that $p(z|\zeta)$ in the subset from which it was 
measured remains accurate in the larger dataset.  

Rossi et al. (2010) have shown that the deconvolution method 
accurately reconstructs the true ${\rm d}N/{\rm d}z$ distribution 
from ${\rm d}{\cal N}/{\rm d}\zeta$.  Figure~\ref{Nzconv} shows that 
the convolution approach also works well, even when only a random 5\% 
of the full dataset is used to calibrate $p(z|\zeta)$ -- as displayed
in Figure \ref{pzzeta}.  Thus, for the 
dataset in which both $z$ and $\zeta$ are available, both the 
convolution and deconvolution approaches are valid, whether or not 
the means (or, for that matter, the most probable values) of $p(z|\zeta)$ 
and $p(\zeta|z)$ are unbiased, and however complicated (skewed, 
multimodal) the shape of these two distributions.  
This remains true in the larger dataset where only $\zeta$ is 
known.  However, whereas the convolution approach assumes that 
$p(z|\zeta)$ is the same in the calibration subset as in the full 
one, the deconvolution approach assumes that $p(\zeta|z)$ is the 
same.  

\subsection{Convolution directly from colors}
\label{cfc}
The integral in equation~(\ref{Nz}) is really a sum over all the 
objects in the photometric dataset, where each object with estimated 
$\zeta$ contributes to ${\rm d}N/{\rm d}z$ with weight $p(z|\zeta)$:
\begin{equation}
 \frac{{\rm d}N(z)}{{\rm d}z} 
  \equiv \int {\rm d}\zeta\,\frac{{\rm d}N(\zeta)}{{\rm d}\zeta}\,p(z|\zeta)
  = \sum_i p(z|\zeta_i).
 \label{Npz|zeta}
\end{equation}
Now, recall that $\zeta$ was the mean (or most probable) value of 
a distribution returned by a photometric redshift code. In cases 
where the observed colours ${\bm c}$ map to a unique value of 
$\zeta$, then this sum over $\zeta$ is really a sum over ${\bm c}$, 
and the expression above is really 
\begin{equation}
 \frac{{\rm d}N(z)}{{\rm d}z} 
  \equiv \int {\rm d}{\bm c}\,\frac{{\rm d}N({\bm c})}{{\rm d}{\bm c}}\,
              p(z|{\bm c}) = \sum_i p(z|{\bm c}_i).
 \label{Npz|c}
\end{equation}
Equation~(\ref{Npz|c}) is one of the key results of this paper.  

\begin{figure*}
\centering
\includegraphics[width=1.0\hsize]{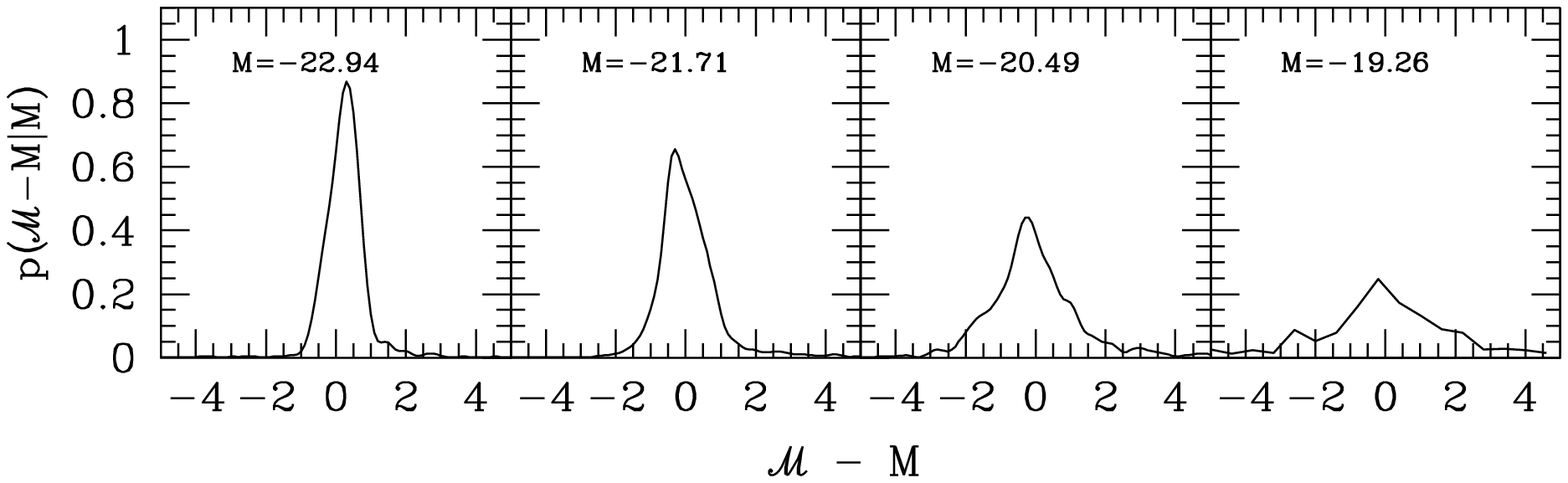}
\includegraphics[width=1.0\hsize]{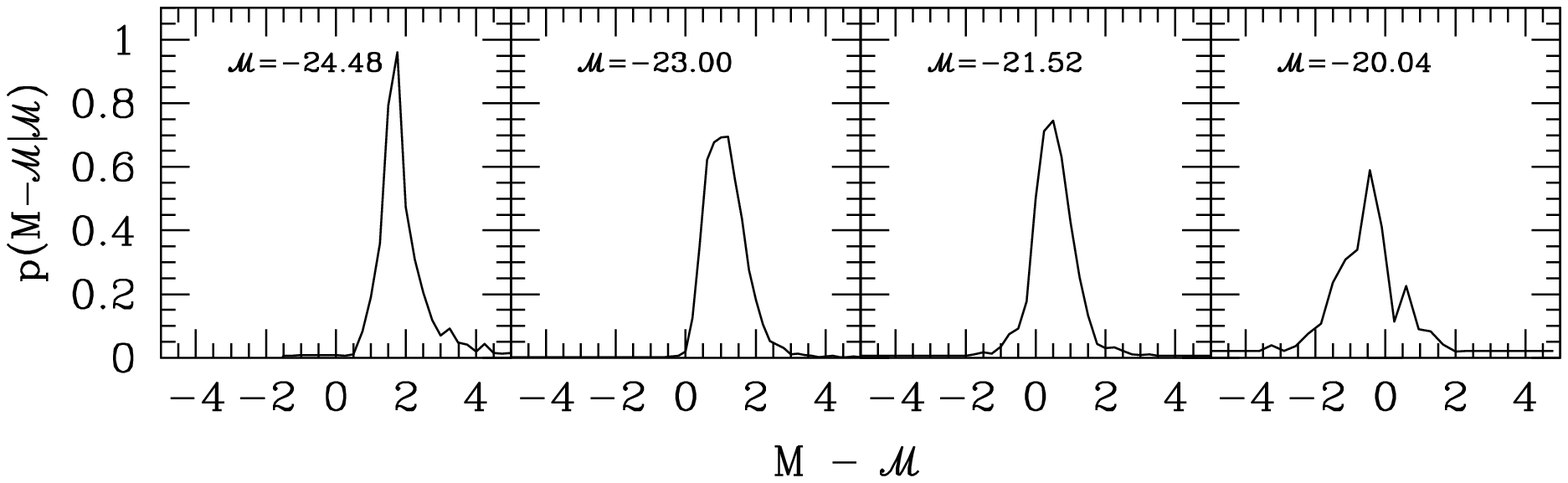}
\caption{Similar to Figure~\ref{pzzeta}, but for the true absolute magnitude 
         and the estimate from the photometry.  Notice that 
         $p({\cal M}|M)$ is approximately symmetrically distributed 
         around $M$, whereas $p(M|{\cal M})$ can be both significantly 
         offset from ${\cal M}$ and skewed.}
\label{MzMphoto}
\end{figure*}

Although we arrived at equation~(\ref{Npz|c}) by requiring the 
mapping ${\bm c}\to \zeta$ be one-to-one (as may be the case for, e.g., 
LRGs), it is actually more general.  This is because one can simply 
measure $p(z|{\bm c})$ in the sample for which spectra are in hand, 
for the same reason that one could measure $p(z|\zeta)$.  
In fact, $p(z|{\bm c})$ is an easier measurement, since it does not 
depend on the output of a photo-$z$ code!  
The constraint on the mapping between ${\bm c}$ and $\zeta$ in 
the discussion above was simply to motivate the connection between 
photo-$z$ codes and the convolution method.  Once the connection has 
been made, however, there is no real reason to go through the 
intermediate step of estimating $\zeta$, since all photo-$z$ codes 
use the observed colors ${\bm c}$ anyway.  In this respect, 
equation~(\ref{Npz|c}) is the more direct and natural expression to 
work with than is equation~(\ref{Npz|zeta}).  In particular, because 
$p(z|{\bm c})$ is an observable, the convolution approach of 
equation~(\ref{Npz|c}) is independent of any photo-$z$ algorithm.  
Of course, if this method is to work, then the subsample with 
spectral information {\em must} be able to provide an accurate 
estimate of $p(z|{\bm c})$.  

\subsection{Relation to photo-$z$ algorithms}\label{pdf}
The convolution method of the previous subsection provides a simple 
way of illustrating how one should use the output from photo-$z$ 
codes that actually provide a properly calibrated probability 
distribution ${\cal L}(z|{\bm c})$ for each set of colors ${\bm c}$,
to estimate ${\rm d}N/{\rm d}z$.  It also shows in what sense the codes should 
be `unbiased'.  

In particular, equation~(\ref{Npz|c}) suggests that one can estimate 
${\rm d}N(z)/{\rm d}z$ by summing over all the objects in the dataset, 
weighting each by its ${\cal L}(z|{\bm c})$.  This is because 
\begin{equation}
 \sum_i {\cal L}(z|{\bm c}_i) = \frac{{\rm d}N(z)}{{\rm d}z} \qquad 
 {\rm if}\  {\cal L}(z|{\bm c}) = p(z|{\bm c}).
 \label{NLz|c}
\end{equation}
Equation~(\ref{NLz|c}) shows that if ${\cal L}(z|{\bm c})$ does not 
have the same shape as $p(z|{\bm c})$, then use of  ${\cal L}(z|{\bm c})$ 
will lead to a bias; this is the pernicious bias which must be 
reduced -- whether or not $\langle z|{\bm c}\rangle$ equals the 
spectroscopic redshift is, in some sense, irrelevant.  
(In the case of a one-to-one mapping between ${\bm c}$ and $\zeta$,
$\langle z|{\bm c}\rangle$ is the same as the quantity 
$\langle z|\zeta\rangle$ which we discussed in the previous subsections.)

Satisfying ${\cal L}(z|{\bm c}) = p(z|{\bm c})$ is nontrivial.  
This is perhaps most easily seen by supposing that the template 
or training set consists of two galaxy types (early- and late-types, 
say), for which the same observed colors are associated with two 
different redshifts.  In this case, if the photo-$z$ algorithms 
are working well, then ${\cal L}(z|{\bm c})$ will be bimodal for at 
least some ${\bm c}$.  However, if the sample of interest only 
contains LRGs, then $p(z|{\bm c})$ may actually be unimodal.  
As a result, ${\cal L}(z|{\bm c}) \ne p(z|{\bm c})$ unless proper 
priors on the templates are used, or care has been taken to insure 
that the training set is representative of the sample of interest.  

\begin{figure*}
\centering
\includegraphics[width=0.88\hsize]{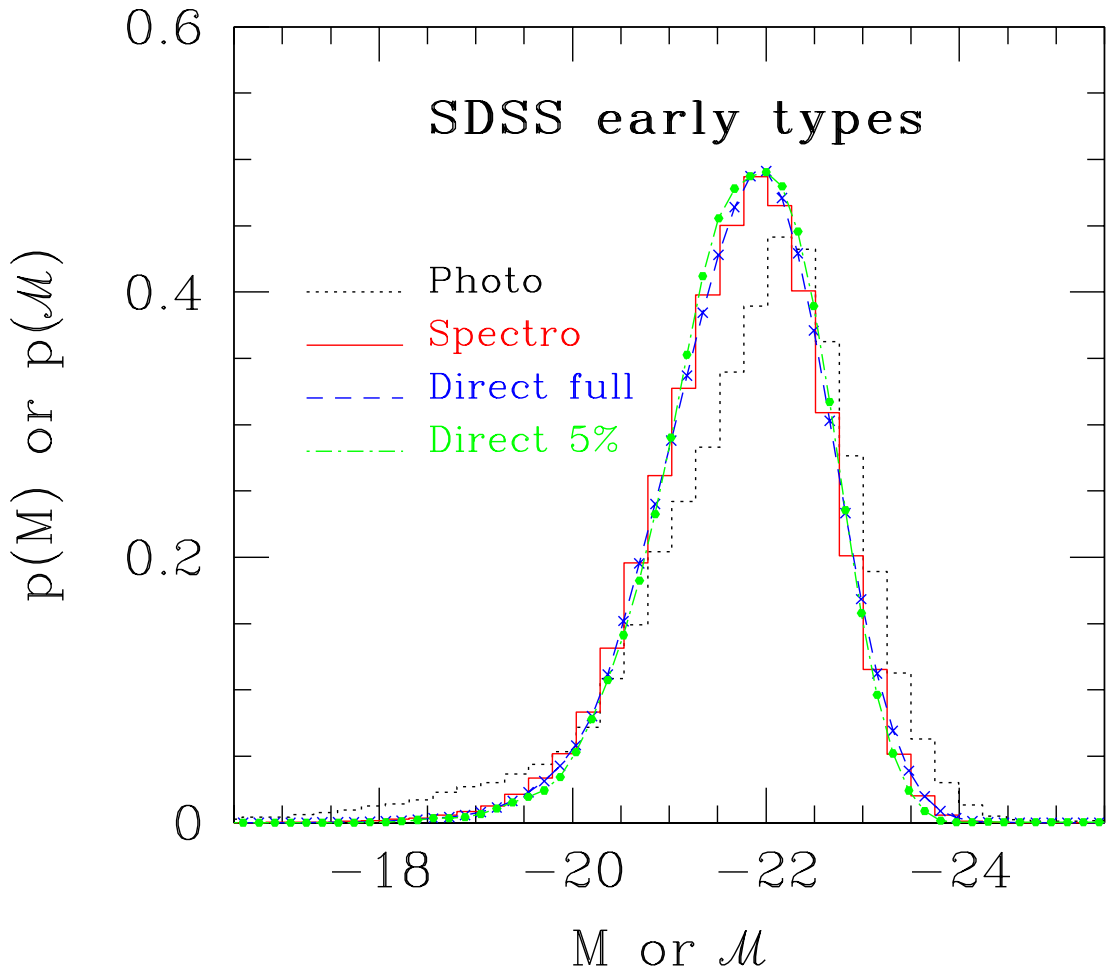}
\caption{Same as Figure~\ref{Nzconv}, but for the absolute magnitudes.
         Crosses show the distribution one obtains by convolving the 
         dotted histogram with the distributions shown in the bottom 
         panel of Figure~\ref{MzMphoto}; solid histogram shows the 
         true distribution of $M$.}
\label{NMconv}
\end{figure*}

\subsection{The luminosity function}\label{phil}
We can perform a similar analysis of the luminosity function.  
In this case, the key is to recognize that, in a magnitude limited 
survey, the quantity which is most directly affected by the 
photometric redshift error is not the luminosity function $\phi(M)$ 
itself, but the luminosity distribution
 $N(M)\equiv V_{\rm max}(M)\,\phi(M)$ (Sheth 2007).  
In a spectroscopic survey, $N(M)$ differs from $\phi(M)$ because 
one sees the brightest objects to larger distances:  $V_{\rm max}(M)$ 
is the largest comoving volume to which an object with absolute 
magnitude $M$ could be seen.  If we use ${\cal M}$ to denote the 
absolute magnitude estimated using the photometric redshift $\zeta$, 
and $M$ its correct value, then 
\begin{equation}
 {\cal N}({\cal M}) = \int {\rm d}M\,N(M)\,p({\cal M}|M).
 \label{NcalM}
\end{equation}
Sheth (2007) describes a deconvolution algorithm for estimating 
$N(M)$ given measurements of ${\cal N}({\cal M})$ and the assumption 
that $p({\cal M}|M)$, measured in a subset for which both $z$ and 
$\zeta$ (hence both $M$ and ${\cal M}$) are available, also applies 
to the full photometric survey.  

Following the discussion in the previous section, we could instead 
have measured $p(M|{\cal M})$, and then used the fact that 
\begin{equation}
 N(M) = \int {\rm d}{\cal M}\,{\cal N}({\cal M})\,p(M|{\cal M})
 \label{NM}
\end{equation}
to estimate the quantity on the left hand side by summing over 
the photometric catalog on the right hand side, weighting each 
object in it by $p(M|{\cal M})$; note that this weight depends 
on ${\cal M}$.  Figure~\ref{MzMphoto} shows $p({\cal M}|M)$ 
and $p(M|{\cal M})$; notice how broad they are, and how much 
more skewed and biased $p(M|{\cal M})$ is than $p({\cal M}|M)$.  
Nevertheless, Rossi et al. (2010) have shown that the deconvolution 
algorithm produces good results.  Figure~\ref{NMconv} shows that 
the convolution algorithm does as well.  

One estimates $\phi(M)$ by dividing $N(M)$ by $V_{\rm max}(M)$.  
Since this weight is the same for all objects with the same $M$, 
one could have added an additional weighting term to the sum above 
to get
\begin{eqnarray}
 \phi(M) &=& \int {\rm d}{\cal M}\,{\cal N}({\cal M})
           \,\frac{p(M|{\cal M})}{V_{\rm max}(M)} \nonumber\\
 &\ne& \int {\rm d}{\cal M}\,\frac{{\cal N}({\cal M})}{V_{\rm max}({\cal M})}
           \,p(M|{\cal M}) .
 \label{phiM}
\end{eqnarray}
One might have written
 $\phi({\cal M}) = {\cal N}({\cal M})/V_{\rm max}({\cal M})$,
so the expression above shows explicitly why the photometric errors 
should be thought of as affecting $N(M)$ and not $\phi(M)$.  

To make the connection to $p(z|{\bm c})$ and then 
${\cal L}(z|{\bm c})$ it is worth considering how one computes 
$M$ from $z$ given the observed colors ${\bm c}$.  If there were 
no $k$-correction, then the luminosity in a given band would be 
determined from the observed apparent brightness by the square of 
the (cosmology dependent) luminosity distance -- the colors are not 
necessary.  
In practice however, one must apply a $k$-correction; this depends 
on the spectral type of the galaxy, and hence on its color.  
As a result, the mapping between $m$ and $M$ depends on $z$ and 
${\bm c}$.  But it is still true that both $M$ and $z$ are 
determined by ${\bm c}$.  Therefore, the spectroscopic subsample 
which was previously used to estimate $p(z|{\bm c})$ also allows
one to estimate $p(M,z|{\bm c})$.  The quantity of interest in the 
previous section, $p(z|{\bm c})$, is simply the integral of 
$p(M,z|{\bm c})$ over all $M$.  The quantity of interest here, 
$p(M|{\bm c})$, is the integral of $p(M,z|{\bm c})$ over all $z$.  
Thus, equation~(\ref{NM}) becomes 
\begin{eqnarray}
 N(M) &=& \int {\rm d}{\bm c}\,\frac{{\rm d}N({\bm c})}{{\rm d}{\bm c}}\,
                           \int {\rm d}z\, p(M,z|{\bm c}) \nonumber\\
&=& \int {\rm d}{\bm c}\,\frac{{\rm d}N({\bm c})}{{\rm d}{\bm c}}\,
                             p(M|{\bm c}) 
 = \sum_i p(M|{\bm c}_i),
\end{eqnarray}
where the second to last expression writes the integral of 
$p(M,z|{\bm c})$ over all $z$ as $p(M|{\bm c})$, 
and the final one writes the integral explicitly as a sum over 
the objects in the catalog.  

The expression above is the convolution-type estimate of $N(M)$; 
it does not require a photometric redshift code.  
However, in principle, a photometric redshift code could output 
${\cal L}(M,z|{\bm c})$: the quantity such codes currently output, 
${\cal L}(z|{\bm c})$, is the integral of ${\cal L}(M,z|{\bm c})$ 
over all $M$.  The relevant weighted sum becomes 
\begin{equation}
 N(M) = \sum_i {\cal L}(M|{\bm c}_i),
\end{equation}
where ${\cal L}(M|{\bm c})$ is the integral of ${\cal L}(M,z|{\bm c})$ 
over all $z$, the sum is over all the objects in the catalog, and 
the method only works if  ${\cal L}(M|{\bm c}) = p(M|{\bm c})$.

Note that the luminosity density (in solar units) can, therefore, 
be written as 
\begin{eqnarray}
 j &\equiv& \int {\rm d}M\,\phi(M)\,10^{-0.4(M-M_\odot)} \nonumber\\
   &=& \int {\rm d}M\,N(M)\,\frac{10^{-0.4(M-M_\odot)}}{V_{\rm max}(M)} 
   \nonumber\\
   &=& \int {\rm d}{\cal M}\,{\cal N}({\cal M})\, 
    \int {\rm d}M\,p(M|{\cal M})\,\frac{10^{-0.4(M-M_\odot)}}{V_{\rm max}(M)}
   \nonumber\\
   &=& \int {\rm d}{\cal M}\,{\cal N}({\cal M})\, 
    \left\langle \frac{10^{-0.4(M-M_\odot)}}{V_{\rm max}(M)}\Biggl|{\cal M}
    \right\rangle \nonumber\\
   &=& \sum_i
    \left\langle \frac{10^{-0.4(M-M_\odot)}}{V_{\rm max}(M)}\Biggl|{\bm c}_i
    \right\rangle .
\end{eqnarray}
The second to last line shows that one requires the average of 
$\langle L/V_{\rm max}(L)\rangle$ summed over the distribution 
$p(M|{\cal M})$; this is easily computed from distributions like 
those shown in the bottom panel of Figure~\ref{MzMphoto}.   
The final expression writes this as a sum over the observed 
distribution of colors.

\subsection{Galaxy scaling relations}\label{phix}
Although the previous section considered the luminosity function in 
a single band, it is clear that the photometric redshift codes could 
output ${\cal L}({\bm M},z|{\bm c})$, where ${\bm M}$ is a set of 
absolute luminosities (typically, these will be those associated 
with the various band passes from which the colors ${\bm c}$ were 
determined).  Hence, the color magnitude relation, which is really 
a statement about the joint distribution in two bands, can be estimated 
by 
\begin{eqnarray}
 N({\bm M}) &=& \int {\rm d}{\bm c}\,\frac{{\rm d}N({\bm c})}{{\rm d}{\bm c}}\,
                      \int {\rm d}z\, p({\bm M},z|{\bm c}) \nonumber\\
&=& \int {\rm d}{\bm c}\,\frac{{\rm d}N({\bm c})}{{\rm d}{\bm c}}\,
                             p({\bm M}|{\bm c}) 
 = \sum_i p({\bm M}|{\bm c}_i).
\end{eqnarray}
Galaxy scaling relations can be estimated similarly, if we simply 
interpret ${\bm M}$ as being the vector of observables which can 
include sizes, etc. (not just luminosities).  In principle, quantities 
other than colors (e.g., apparent magnitudes, surface brightness, 
axis ratios) can play a role in the photometric redshift determination; 
this can be incorporated into the formalism simply by using ${\bm c}$ 
to now denote the full set of observables from which the redshift and 
other intrisic quantities ${\bm M}$ were estimated.  

If one wishes to use the output from a photo-$z$ code, rather than 
from the spectroscopic subset, one would use  
\begin{eqnarray}
 N({\bm M}) = \sum_i {\cal L}({\bm M}|{\bm c}_i),
\end{eqnarray}
having checked that, in the spectroscopic subset, 
 ${\cal L}({\bm M}|{\bm c}_i) = p({\bm M}|{\bm c}_i)$.

\section{Discussion}
We showed how previous work on deconvolution algorithms for making 
unbiased reconstructions of galaxy distributions and scaling 
relations (Sheth 2007; Rossi \& Sheth 2008; Rossi et al. 2010) 
could be related to convolution-based methods.  
Whereas deconvolution based methods require accurate knowledge of 
$p(\zeta|z)$, the distribution of the photometric redshift $\zeta$ 
given the true redshift $z$, convolution based methods require accurate 
knowledge of $p(z|\zeta)$.  Since $\zeta$ is derived from photometry, 
this may more generally be written as $p(z|{\bm c})$, where ${\bm c}$ 
is the vector of observed photometric parameters which were used to 
estimate the redshift.  In both cases, $p(z|{\bm c})$ and $p(\zeta|z)$ 
are calibrated from a sample in which $z$ is known, and are then 
used in a larger sample where $z$ is not available.  If the smaller
training set has the same selection limits as the larger dataset 
(e.g., both have the same magnitude limit) then both approaches are 
valid. 	We illustrated our arguments with measurements in the SDSS
(Figures~\ref{pzzeta}--\ref{NMconv}).  

We also showed what additional information must be output from 
photometric redshift codes if their results are to be used in a 
convolution-like approach to provide unbiased estimates of galaxy 
scaling relations.  In particular, we argued that only if the redshift 
distribution output by a photo-$z$ algorithm, ${\cal L}(z|{\bm c})$, 
has the same shape as $p(z|{\bm c})$, can the algorithm be said 
to be unbiased.  Only in this case its output (available for 
the full sample) can be used in place of $p(z|{\bm c})$ (which is 
typically available for a small subset).  The safest way to accomplish 
this is for the training set to be a random subsample of the full 
dataset -- and to then tune the algorithm so that 
${\cal L}(z|{\bm c}) = p(z|{\bm c})$.  
If the training set is not representative, then care must be taken 
to ensure that ${\cal L}(z|{\bm c})$ does not yield biased results.  

Obtaining spectra is expensive, so the question arises as to whether 
or not there is a more efficient alternative to the random sample 
approach.  For the convolution method, which requires 
$p(z|{\bm c})$, the answer is clearly `yes'.  This is because some 
color combinations (e.g. the red sequence) might give rise to a 
narrow $p(z|{\bm c})$ distribution, whereas others may result in 
broader distributions.  Since it will take fewer objects to 
accurately estimate the shape of a narrow $p(z|{\bm c})$ distribution 
than a broad one, observational effort would be better placed in 
obtaining spectra for those objects which produce broad 
$p(z|{\bm c})$ distributions.  
For the deconvolution approach, one would like to preferentially 
target those redshifts $z$ which produce broader $p(\zeta|z)$ 
distributions -- for similar reasons.  But, since $z$ is not known 
until the spectra are taken, this cannot be done, so taking a 
random sample of the full dataset is the safest way to proceed.  

Our methods permit accurate measurement of many scaling relations 
for which spectra were previously thought to be necessary (e.g. 
the color-magnitude relation, the size-surface brightness relation, 
the Photometric Fundamental Plane), so we hope that our work will 
permit photometric redshift surveys to provide more stringent 
constraints on galaxy formation models at a fraction of the cost of 
spectroscopic surveys.  


\section*{Acknowledgments}
RKS thanks L. Da Costa, M. Maia, P. Pellegrini, M. Makler and 
the organizers of the DES Workshop in Rio in May 2009 where he 
had stimulating discussions with C. Cunha and M. Lima about the 
relative merits of convolution and deconvolution methods, 
and the APC at Paris 7 Diderot and MPI-Astronomie Heidelberg, 
for hospitality when this work was written up.

Funding for the SDSS and SDSS-II has been provided by the Alfred P. 
Sloan Foundation, the Participating Institutions, the National Science 
Foundation, the U.S. Department of Energy, the National Aeronautics and
 Space Administration, the Japanese Monbukagakusho, the Max Planck 
Society, and the Higher Education Funding Council for England. The 
SDSS Web Site is {\tt http://www.sdss.org/}.

The SDSS is managed by the Astrophysical Research Consortium for the
Participating Institutions. The Participating Institutions are the
American Museum of Natural History, Astrophysical Institute Potsdam,
University of Basel, University of Cambridge, Case Western Reserve
University, University of Chicago, Drexel University, Fermilab, the
Institute for Advanced Study, the Japan Participation Group, Johns
Hopkins University, the Joint Institute for Nuclear Astrophysics,
the Kavli Institute for Particle Astrophysics and Cosmology, the 
Korean Scientist Group, the Chinese Academy of Sciences (LAMOST), 
Los Alamos National Laboratory, the Max-Planck-Institute for Astronomy 
(MPIA), the Max-Planck-Institute for Astrophysics (MPA), New Mexico 
State University, Ohio State University, University of Pittsburgh, 
University of Portsmouth, Princeton University, the United States 
Naval Observatory, and the University of Washington. 



\label{lastpage}

\end{document}